# A Note on Semantics
## (with an Emphasis on UML)

Bernhard Rumpe
Institut für Informatik
Technische Universität München
D- 80290 Munich, Germany
Bernhard.Rumpe@in.tum.de, www.in.tum.de/~rumpe

> "In software engineering people
> often believe a state is a node
> in a graph and don't even care
> about what a state means in reality."
> David Parnas, 1998

**Abstract**

This note clarifies the concept of *syntax* and *semantics* and their relationships. Today, a lot of confusion arises from the fact that the word "semantics" is used in different meanings. We discuss a general approach at defining semantics that is feasible for both textual and diagrammatic notations and discuss this approach using an example formalization. The formalization of hierarchical Mealy automata and their semantics definition using input/output behaviors allows us to define a specification, as well as an implementation semantics. Finally, a classification of different approaches that fit in this framework is given. This classification may also serve as guideline when defining a semantics for a new language.

## 1  Introduction

With the standardization of the Unified Modeling Language (UML) [1], there is currently an ongoing, vivid discussion about the precise semantics of its constructs. Whereas the OMG was responsible for the standardization of the UML regarded as a notation, the semantics is still under investigation. There is a large amount of theoretical work available that discusses subsets and adaptations of the UML, see e.g. [31], by giving it a precise semantics and extracting results from that. This widely excellent work, however, needs itself some investigation, as often some implicit assumptions are made, which influence the way a formalization is defined and the results of such a formalization. In particular, it is difficult to compare the results of the respective articles, as this comparison is based on the treated subset of the notation, the assumptions on the kind of developed systems, the relationship defined between the new constructs and the given notion of systems, and, furthermore, on the way all this is being written down (formalized).

Therefore, in this paper, we give a general framework of how to define semantics of a notation and classify different approaches within this framework. We discuss different variants of semantics definitions their goals, benefits and drawbacks. Although our work is not particularly dedicated to UML, the work we refer to mainly stems from approaches at formalizing UML.

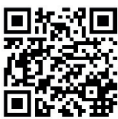

The paper is organized as follows. In the next section we clarify the concept of "syntax" and how syntax is represented. Section 3 focuses on the concept of "semantics", introducing the semantic domain and the semantics mapping. In Section 4, an example semantics definition is given, which shows some of the possible properties a semantics definition may exhibit. Section 5 gives a classification of different variants of semantics definitions that can be found in the literature. This classification may as well serve as guideline for semantics definitions. Section 6 discusses the future of semantics definitions for UML. Section 7 concludes.

## 2 Syntax

The UML documents [1] contain a paper called the "Semantics of UML". However, this paper does not only focus on semantics but mainly on syntactic issues. There are informal insights into the semantics of UML given in this semantics paper. This is sufficient for experienced users to gain more insights into the purpose of the constructs of UML. However, when details are considered lots of ambiguities remain. This has been shown by a whole load of papers dealing with packages [3], Class Diagrams [27] [28], State Diagrams [5] [10], or even their integration [26] [29].

The term "syntax" is used whenever we refer to some notation. In conventional textual notations, syntax is described by the set of characters used (alphabet) and their possible sequences (symbols, strings, sentences). We refer to the set of all correct sequences as a *language*. When diagrams are involved, the situation becomes somewhat more complex: the syntax does not deal with the linear sequence of possible characters anymore, but must cope with two- or even three dimensional diagrammatic elements, like different kinds of boxes (denoting states, classes, objects, etc.) and different kinds of arcs (denoting transitions, associations, links, etc.).

Syntactical issues purely focus on the notation, completely disregarding any intention behind the notation. In principle, the syntax defines a language $\mathcal{L}$ of well formed declarations and statements. If we again have a look at conventional textual languages, we find several layers of defining such a language.

1. A set of characters is defined to form an alphabet.
2. The characters are grouped into words, denoting key words, names, numbers, delimiters, etc. This lexical layer typically uses regular expressions.
3. A third layer groups these words into sentences using a context free grammar.
4. A fourth and final layer constrains these sentences by giving context conditions.

If diagrams are involved, the situation becomes somewhat more difficult. Today it is still not clear what the best notation is that allows us to describe these graphical parts. However, compared with text, we can assume that simple lines and characters correspond to the alphabet. Several kinds of boxes and arcs then correspond to the lexical layer. Boxes and arcs are given different shapes to denote different kinds and both are parameterized with textual attributes, e.g. class or method names or even expressions, which themselves can be defined using a textual grammar.

In UML the third layer, normally using a context free grammar, is replaced by the meta-model. Although there are graph grammars available [4] that extend the textual ideas to diagrams, and have been applied to significant parts of UML already [3] [5], for practical purposes it is feasible to use Class Diagrams to model the abstract syntax (meta-model) of UML. Class Diagrams on the one hand are a notation, well known by the UML users, and on the other hand a notation well suited for implementation by tool vendors. However, when used for a semantics definition, the notation by no means should be defined recursively using itself. Instead, a notation should be used that already has a semantics.

In UML some context conditions are partly given using the Object Constraint Language (OCL), which is also part of UML, and partly stated in English. In textual languages, these constraints can be stated more precisely, e.g. by attribution of the abstract syntax tree, which results from the parsing through the context free grammar. Table 1 contains a comparison of the most common techniques for defining syntax today.

**Table 1: Describing textual and diagrammatic notations**

| **Layer** | **Textual Notation** | **Diagrammatic Notation** |
| --- | --- | --- |
| Alphabet | Set of characters | (Lines, characters) |
| Lexical Syntax | Words | Set of icons (boxes, arcs, etc) |
| Context free syntax | Context free grammar, Result: abstract syntax tree | Meta-model (Class Diagram) Result: "model" |
| Context conditions | Attribution of abstract syntax tree | OCL |

The meta-model of UML gives a precise notion of what the abstract syntax is. However, it does not cope with semantics. Furthermore, context conditions are by no means "semantic conditions", but purely constrain the syntax. They give well formedness rules, e.g. each variable must be defined before used, without telling you what a variable is. Or can the semantics of C++ be understood from the context free grammar and context conditions like well-formed typing (without knowledge of similarly structured languages)? Of course, context conditions aim at a resulting language to which a sound semantics can be given. Therefore, it is quite common to term context conditions also "semantic conditions".

The different layers of defining syntax allow at each layer to cope with the issues important at that layer. The meta-model and the abstract syntax tree, as well as the latter's more efficient implementation the directed acyclic graphs (DAG), all three serve to define the abstract syntax of some language. There are some fundamental differences, similar to those between the functional and the object-oriented paradigm. Abstract syntax trees are hierarchical and therefore allow to think in a purely decomposition oriented manner. Meta-models instead are relation-based and therefore do not have a canonical point, where to start with the semantic definition by decomposing a hierarchy into smaller parts.

As we see from the above discussion, the definition of a language $\mathcal{L}$ itself needs the use of a notation, which we for now call $\mathcal{N}_\mathcal{L}$. For textual languages, $\mathcal{N}_\mathcal{L}$ is typically a combination of the Backus Naur Form (BNF) and Chomsky-2 context free grammars (CH-2). In the UML case, this notation is a combination of Class Diagrams and OCL used to define the meta-model.

The use of a notation $\mathcal{N}_\mathcal{L}$ to define the language $\mathcal{L}$ does not only result in the definition of $\mathcal{L}$, but usually also gives an abstract version of the language, the abstract syntax tree or the meta-model, and often also gives an algorithm (program) how to transform (parse) the language $\mathcal{L}$. For now let us identify the language and its abstract version and discuss how to proceed when defining semantics.

# 3 Semantics

A semantics definition consists of two parts: first, a semantics domain is to be defined, then, a mapping from the syntax to the semantics domain is to be provided. The following two sections discuss these two parts.

## 3.1 Semantic Domain

The semantics of a language tells us about the *meaning* of each construct of the language $\mathcal{L}$ in question. This is usually done by explaining new constructs of the language in terms of already known (and hopefully well understood) concepts. This situation does not only occur when semantics is defined formally, but also when semantics is explained informally. Thus a certain domain is needed that is assumed to be a well understood set of concepts. We call this domain the *semantic domain* and denote it with $\mathcal{S}$. In the UML the description of the semantic domain contains objects, values, messages that are passed around, etc. Unfortunately, this is informally explained and scattered almost throughout the whole UML description [1]. Indeed it is not uncommon in papers that introduce a new notation to give at least an informal notion of this semantic domain.

The semantic domain $\mathcal{S}$ also gives a notion of what concepts exist in the universe of discourse. This can be compared with the UoD (universe of discourse) of RM-ODP [35], where also the system of concern is described. In [6] we have explicitly defined such a semantic domain and called it "system model", as it describes our notion of what a system is. Therefore, the semantic domain is an abstraction of reality, describing the important aspects of the systems that we are interested in to develop.

The explicit definition of the semantic domain is a very important and crucial issue, as it on the one hand allows us to understand the intended kind of systems a notation is designed for, and on the other hand is a prerequisite to compare different semantics definitions.

As it was with the syntactic case, we need an underlying notation $\mathcal{N}_s$ that allows us to describe the semantic domain $\mathcal{S}$. The variety of notations used for this purpose is much larger than in the syntactic case. As candidates almost all general purpose formal languages, like Z [7], algebraic specification languages [8], [9] or pure mathematics apply. We will discuss this variety and the implications of such a choice later in greater detail.

In practice semantics and syntax are often mixed up leading to considerable confusion. While syntax is the notation and therefore is what we manipulate, semantics tells us about the meaning of the notation. We do not directly manipulate semantics. Instead everything we see and work with on the paper or on the screen is a syntactic representation. Unfortunately this means that if we define "semantics" we also need a syntactical representation of it. As we will see below, it is therefore to some extent dependent on the purpose of a language to regard it as syntax or as semantics.

One common misconception is to talk about semantics, when "behavior" is meant. Behavior as opposed to structure of a system are both views of the system, and both are represented e.g. in UML using syntactic concepts.

## 3.2 Semantic Mapping

Given both, a new notation $\mathcal{L}$ and a semantic domain $\mathcal{S}$ the third step of a semantics definition is to relate the syntactic concepts with the concepts of the semantic domain. When semantics is explained informally, then this is usually done by explaining which new concept maps to which given one. A good example for that are associations in Class Diagrams of

UML. It is common to explain an association by its implementation, either as a set of links between the attached classes, or (in more complex cases) as a class on its own. In both cases the implementation must obey the invariant imposed by the implemented association. In principle, the semantics definition is a mapping

$$\mathcal{M} : \mathcal{L} \to \mathcal{S}$$

Here, we disregard that in general a document imports others and therefore its semantics depends on the semantics of the imported ones. This can e.g. be incorporated using a mapping $\mathcal{M}$ that takes an interpretation of imported symbols as parameter.

Based on the different kinds of notations used to describe the syntax ($\mathcal{N}_\mathcal{S}$) and the semantics ($\mathcal{N}_\mathcal{L}$) there is a real variety of semantics mappings, which we will also discuss in detail later. In many "formalizations", this mapping is given informally, as it is explained by mapping examples from $\mathcal{L}$ to $\mathcal{S}$. The mapping $\mathcal{M}$ itself is not explicitly given. However, if $\mathcal{M}$ would be given explicitly, a notation would be required to describe it. So there is another notation $\mathcal{N}_\mathcal{M}$ that is used to describe the mapping. Despite the variety of notations for syntax and semantics, there are only few notations used for the mapping. On the one hand mathematics [10] [11] [12] [13], and on the other hand graph transformations [3] [5] are used.

Interestingly there seems to be no approach that uses Z, an algebraic specification language or something similar to explicitly define such a mapping. This might be partly due to the fact that the notation for the mapping must include the notations for the syntactic and the semantic domains ($\mathcal{N}_\mathcal{L}, \mathcal{N}_\mathcal{S} \subseteq \mathcal{N}_\mathcal{M}$). This works for graph transformations if both domains are graph structures, and it works for mathematics, as everything can be embedded here, but it needs major work to model the syntactic domain (which is essentially a context free language or a graph) within Z or something similar.

The explicit definition of the mapping $\mathcal{M}$ allows us to reason about it. Today there are approaches that define $\mathcal{M}$ in an algorithmic fashion, and even implement it. Their idea is to provide this mapping to the software engineer. This enables the software engineer to translate documents of the informal language $\mathcal{L}$ into documents of the formal notation $\mathcal{S}$ and use proof and analysis techniques on $\mathcal{S}$. For example let us assume there is a predicate "consistent: $\mathcal{S} \to$ Bool", describing that a proper implementation of a document written in $\mathcal{S}$ can be found. A prerequisite for this property is that no contradiction can be found within that document. The software engineer can apply it to the transformed documents. It is a drawback of this approach that the engineer must be capable and willing to understand not only the notation $\mathcal{N}_\mathcal{L}$, but also the notation $\mathcal{N}_\mathcal{S}$. Typically the engineer is not really interested in the semantics notation $\mathcal{N}_\mathcal{S}$, but only wants to cope with $\mathcal{N}_\mathcal{L}$. Therefore, a better approach would emerge if the definer of the semantics would prove that for all documents $d \in \mathcal{L}$ the resulting semantics is consistent:

$$\forall d \in \mathcal{L}. \ \text{consistent}(\mathcal{M}(d))$$

Then the software engineer using the language $\mathcal{L}$ could be sure to have consistent models without being explicitly exposed to the formally given semantics domain, but only needs an informal "feeling" about the semantics domain in order to successfully deal with the provided techniques and tools. If such a proof of consistency for $\mathcal{L}$ cannot be provided by the method developer, context conditions constraining the notation $\mathcal{L}$ such that the above condition becomes true should be introduced. Precisely: context conditions must be found that are defined purely on the syntax and that are sufficient to ensure consistency. So $\mathcal{L}$ should be restricted to the set { $d \in \mathcal{L}$ | consistent($\mathcal{M}(d)$) } of consistent documents.

It is one of the most important positive results that a formalization of a notation can have, if it induces an improved version of the notation. Not the formalization itself is of much help, but the insights concluded from it are. These insights mostly come from analyzing the formalization according to the following issues, and are primarily drawn by the same persons that gave the formalization (ideally these are definers of the notation) and not by the users of a notation:
- Does the given formalization capture the intuition of the users?
- Are the context conditions sufficient to ensure consistency?
- Does the notation allow to capture important properties of the semantic domain?
- Are transformation rules etc. sound wrt. the given semantics?

It probably needs a tremendous amount of work to capture these and related questions, but it surely must be done. A necessary prerequisite at least for success wrt. the last purpose is the explicit definition of the semantics mapping $\mathcal{M}$. Other questions, like the first one, also aim at a consensus between users. This can only be achieved by a broadly accepted standardization, based on a formalization of syntax and semantics.

## 4 Example Semantics Definition

In this section, we will illustrate the principles of semantics by means of a small example. It is mainly the way we defined the example that is of interest. To make the example not too simple, we use hierarchical Mealy automata (HMA). These are automata that exhibit a hierarchically decomposed state concept and their transitions are each labeled with one input and one output character. Transitions may cross state borders, nondeterminism and partiality are allowed.

In a first step we formalize HMA, then we define an appropriate semantics domain and finally we give the semantics mapping. To define these three parts, we use one underlying notation for all three: plain mathematics is used for $\mathcal{N}_\mathcal{M}$, $\mathcal{N}_\mathcal{S}$, and $\mathcal{N}_\mathcal{L}$. As the mapping $\mathcal{M}$ becomes somewhat complex, we construct it in three layers, which are finally composed using functional composition. We also introduce two different kinds of semantics, one suited for specification and one suited for implementation.

### 4.1 Hierarchical Mealy Automata

The syntax of HMA is defined using a tuple construct. An HMA is mathematically given by a tuple $(S,\leq,M,\delta,I)$, where
- S is a set of states,
- $(\leq) \subseteq S \times S$ is a non-reflexive, transitive containment relation for states,
- M is a set of messages,
- $\delta \subseteq S \times M \times M^\varepsilon \times S$ is a transition relation, and
- $I \subseteq S$ is a set of initial states.

By $\mathcal{HMA}$ we denote the set of such tuples. Thus $\mathcal{HMA}$ serves as our syntactic domain $\mathcal{L}$. By $M^\varepsilon$ we denote the set M extended by the special element $\varepsilon$. The transition relation relates source state, input message, an optional output message and the destination state. The omission of an output message is denoted by using $\varepsilon$. A simple but typical HMA is given in Figure 1.

Please note that we have introduced the context condition that $\leq$ is a partial ordering. So far we have neither demanded that the set I is not empty, nor that $\leq$ is a hierarchy (disallowing overlapping states), nor that any set is finite.

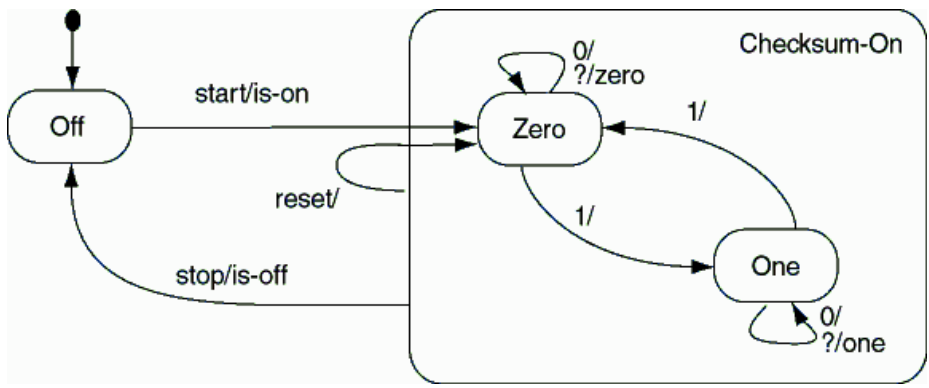

**Figure 1: An HMA for Checksum Calculation**

### 4.2 Semantic Domain for Hierarchical Mealy Automata

When regarding the semantics of a notation, we first decide which kinds of systems we are interested in. The more complex a notation usually is, the more specific is its application area and the less different kinds of systems it is suited for. In our simple case, it is sufficient to decide we have objects, and these objects communicate with each other via message passing. For each incoming message, an object is allowed to send at most one answer. As the syntax we defined does not include a composition operator, we do not need to talk about object interaction (asynchronous, synchronous, in-parallel, etc.), we only assume that objects may not reject incoming messages (as it is the case in sequential languages). We regard the state of objects to be encapsulated. Therefore it is a good choice to take the external visible behavior as semantics. Neither the state nor the transition relation are visible in the semantics. We define this behavior in terms of a relation between the sequence of inputs and the sequence of outputs. Defining M* as the sequence of messages over M, then our semantics $\mathcal{S}$ is the set of pairs of input and output sequences, defined as $\wp(M^* \times M^*)$. Thus the mapping

$$\mathcal{M} : \mathcal{HMA} \rightarrow \wp(M^* \times M^*)$$

is the goal of our semantics definition. Please note that in the semantics states do not show up. This reflects that states are an encapsulated within objects and therefore can not be observed from outside.

The semantics of HMA is given in three layers. This is due to the fact that a one-step semantics definition is quite involved. In order to manage the semantics definition, we introduce two intermediate stages, one including a new formalism (which is only a slightly changed version of HMA), and another, where we cope with only a substyle of the syntax. It is important that we carefully motivate our decisions.

### 4.3 Transition to Flat Mealy Automata

The first step is done by removing the hierarchy within automata. So basically the resulting flat Mealy automata (MA) are a subset of the HMA, disregarding the containment ≤. The set of Mealy automata $\mathcal{MA}$ is therefore described as the set of tuples $(S_2, M_2, \delta_2, I_2)$ and serves as the semantics in this first layer of semantics definition. The motivation behind the reduction transformation comes from the fact that in the final semantics states do not show up. Thus it is ok to flatten the state hierarchy as long as the transition structure is preserved.

We assume that the nesting of states in HMA indicates that if an object is in a substate, it is also in its according superstates. Thus if the object is in state $s \in S$, it in fact is in each of the states $\{t \in S \mid s \leq t\}$. Furthermore, if a state is decomposed into a set of substates, we demand that an object in this state must be in exactly one of the substates. Therefore, substates "partition" the superstate.

The new states $S_2$ therefore can be defined as linearly ordered subsets of $S$ that are upwards and downwards closed. Interestingly, if we assume that $\leq$ does not have infinite descending chains, each of these sets over $S$ corresponds to exactly one set of basic states. As this requirement is in practice not a hard constraint on HMAs, we accept this constraint and identify each such set with its smallest element. So we end up with the following translation:

$S_2 = \{ s \in S \mid \neg \exists\, t \in S.\ t < s \}$

$M_2 = M$

$\delta_2 = \{ (s_2,i,o,t_2) \in S_2 \times M \times M^\varepsilon \times S_2 \mid \exists s,t \in S.\ (s,i,o,t) \in \delta \wedge s_2 \leq s \wedge t_2 \leq t \}$

$I_2 = \{ s_2 \in S_2 \mid \exists s \in I.\ s_2 \leq s \}$

To complete the mapping definition, we define the first part of $\mathcal{M}$:

$\mathcal{M}_1 : \mathcal{HMA} \to \mathcal{MA}$

$\mathcal{M}_1(S,\leq,M,\delta,I) = (S_2,M_2,\delta_2,I_2)$

Observe, that in case $S$, $I$ and $\delta$ are finite, this transformation can easily be implemented. A concrete program could do this transformation, and graph transformations can do exactly the same [4].

### *4.4 Transition Completion*

Because we decided that the object may not reject any incoming message, the relation that we regard as behavior description of an object must therefore be complete in the sense that it offers at least one possible reaction for each incoming input sequence. If we were to immediately proceed with the next step of semantics mapping, we would find out that the result of the semantics mapping does not satisfy this requirement.

This has two reasons: on the one hand, if the set of initial states is empty, then the result is also empty (see semantics definition below). As a result of this we impose the context condition for MA (and hence similarly for HMA) that $I \neq \emptyset$ which seems a good constraint from a methodical point of view.

The other problem is due to the fact that we allow the transition relation $\delta$ to be partial. An MA is partial for a given state $s \in S_2$ and input $i \in M$ if it doesn't contain a transition accepting input i. For HMA this means that a state $s \in S$ is partial in input $i \in M$ if neither state s nor one of its superstates and not all of its substates contain a transition accepting input i. We can interpret this in different ways. Among them are three major possibilities:

- Ignore: The incoming message is just ignored, no state change and no output occurs.
- Chaos: The incoming message was not awaited, everything, including a crash may happen.
- Underspecification: The specifier did say nothing about possible behavior. Nothing is forbidden, everything is allowed.

Interestingly the latter two interpretations lead to the same semantics, which roughly corresponds to the loose approach used in algebraic specification techniques [11] [8] and is suitable when the HMA are interpreted as a specification technique. The former instead is useful if HMA are regarded as a programming notation and leads to a default implementation, where the HMA was partial. This roughly corresponds to the notion of initial semantics. Giving one notation two different semantics in different contexts of usage is quite useful and has been done e.g. in [13]. We therefore define two variants of semantics, respecting these two interpretations in different contexts. It is most easily done by adding appropriate transitions to the transition relation to complete it. The implementational semantics:

$\mathcal{M}_{2i}: \mathcal{MA} \to \mathcal{MA}$

$\mathcal{M}_{2i} (S,M,\delta_2,I) = (S,M,\delta_{3i},I)$, where

$\delta_{3i} = \delta_2 \cup \{ (s,i,\varepsilon,s) \mid s \in S, i \in M, \neg\exists o \in M^\varepsilon, t \in S. (s,i,o,t) \in \delta \}$

and the specification semantics:

$\mathcal{M}_{2s}: \mathcal{MA} \to \mathcal{MA}$

$\mathcal{M}_{2s} (S,M,\delta_2,I) = (S,M,\delta_{3s},I)$, where

$\delta_{3s} = \delta_2 \cup \{ (s,i,u,p) \mid s,p \in S, i \in M, u \in M^\varepsilon, \neg\exists o \in M^\varepsilon, t \in S. (s,i,o,t) \in \delta \}$

Please note, that both mappings lead to a complete MA, and both mappings do not affect the behavior when explicitly given by transitions.

## 4.5 Behavioral Semantics

From the completed flat Mealy machines, it is now quite easy to map to possible behaviors. The third part of the mapping $\mathcal{M}$ is given by the translation:

$\mathcal{M}_3 : \mathcal{MA} \to \wp(M^* \times M^*)$

This basically requires to mimic the step by step transition semantics of an automaton through a recursive definition. The idea is that if we have given the behavioral semantics of the automaton for n steps, we can define the behavior for (n+1) steps by performing a single step and then perform n steps. As we change states in each step, we use the actual state we start with as an additional parameter.

Assuming $R_n \in S \to \wp(M^* \times M^*)$ is a state-parameterized relation that describes the possible behavior for the first n steps. Regarding that (i++iq) denotes the attachment of element i to a sequence iq and $\varepsilon$ is the empty sequence ($\varepsilon$++iq=iq), then $R_{n+1}$ is defined by:

$R_{n+1}(s) = \{ (i\text{++}iq, o\text{++}oq) \mid \exists t. (s,i,o,t) \in \delta \land (iq,oq) \in R_n(t) \}$

With the initial $R_0$ being defined as the relation that contains exactly the behavior if an empty input sequence arrives:

$R_0(s) = \{ (\varepsilon,\varepsilon) \}$

we can finally define $\mathcal{M}_3$ using the union of all such relations wrt. the given initial states:

$\mathcal{M}_3 (S,M,\delta,I) = \cup_{n \in \mathbb{N}, s \in I} R_n (s)$

The behavior of a complete flat Mealy automaton is the union of all behavior relations that are explicitly described by the MA.

The overall semantics mapping for HMA is then given by the functional composition and comes in the implementational and the specificational variant:

$\mathcal{M}_i = \mathcal{M}_3 \circ \mathcal{M}_{2i} \circ \mathcal{M}_1$

$\mathcal{M}_s = \mathcal{M}_3 \circ \mathcal{M}_{2s} \circ \mathcal{M}_1$

## 4.6 Results

We have already collected some important results while carrying out the definition. Kinds of results are:

- Context conditions which have been introduced to get a sound semantics, and which usually also prove sensible from a methodical point of view. A trivial example was the introduction of I≠∅. A less trivial, but also suitable context conditions is the restriction to only finitely nested states.

- Insights and clarifications of the meaning of concepts. Rather interesting was the discovery of the two different versions of treatment of partiality.

Furthermore, when introducing these steps towards semantics, we have taken some (quite reasonable and well motivated) assumptions. If reject this assumptions, then the given semantics can be essentially wrong. However, our assumptions are given explicitly such that we can recognize and argue about them. By just reading a formalization, usually the properties of the definition do not become apparent (except the reader is an expert who has seen several similar definitions). So the formalization must be analyzed in order to increase confidence in that the formalization is correct and meets our intuition. The two main techniques for that are testing examples and proving properties. This is quite similar to analyzing a given UML model versus the intuition. We will just state some results here without providing the formal proofs (see [11] for similar proofs on a more complex version or do examples on your own). Some statements:

- A deterministic and complete HMA leads to a deterministic behavior relation (one possible output sequence for each input).
- Both completion transformations lead to complete MA and therefore both semantic mappings $\mathcal{M}_i$ and $\mathcal{M}_s$ are well defined.
- The ignoring semantics is a subset of the underspecification semantics: for all hierarchic Mealy automata A we have $\mathcal{M}_i(A) \subseteq \mathcal{M}_s(A)$.
- The semantics definition resembles erratic choice of a possible transition for each input message. Behavior completion after the mapping into a relation would result in a substantially different semantics as the nondeterminism in the MA would be interpreted angelic: An implementation must follow all possible paths staying non-chaotic as long as possible. The angelic (and also the demonic) interpretation of nondeterminism are to some extent interesting semantics. Unfortunately, both lead to very complex implementations and are hard to refine as this kind of nondeterminism does not correspond to underspecification.

Besides these useful insights, which partly help the notation developer to increase confidence in the notation and partly can also help the user of the notation, a next step for the notation developer now should be to define constructive manipulation techniques on the notation.

In our case e.g. techniques for inheriting HMA between classes were of interest. Inheritance of behavior allows us to substitute an object by a new, refined one. Fortunately we do have a good notion of behavioral refinement, defined in [14]. It is simply based on set inclusion. So we are looking for techniques that allow to constructively modify HMA $A_1$ into $A_2$ such that

$$\mathcal{M}_s(A_2) \subseteq \mathcal{M}_s(A_1)$$

holds. For the implementational semantics this relation is not necessary, as we do not refine implementations, but we do refine specifications.

In Figure 2 the above formula is depicted in a commuting diagram. Lots of similar techniques have been defined and quite successfully used for structural notations, namely the normalization guidelines in relational database design [15] or re-factoring functionality in class diagrams [16]. Also a refinement calculus for object structures has been defined in [17]. And indeed we can find lots of rather simple rules that allow us to constructively modify a HMA document $A_1$ into $A_2$ and that ensure the formula given above. Almost all such rules come along with appropriate context conditions. Among them are quite simple rules like removing of initial states as long as at least one exists, removing unreachable states, more complex ones like splitting a state into several ones in order to allow a more detailed behavior description, and intrusion of a state in another nested one.

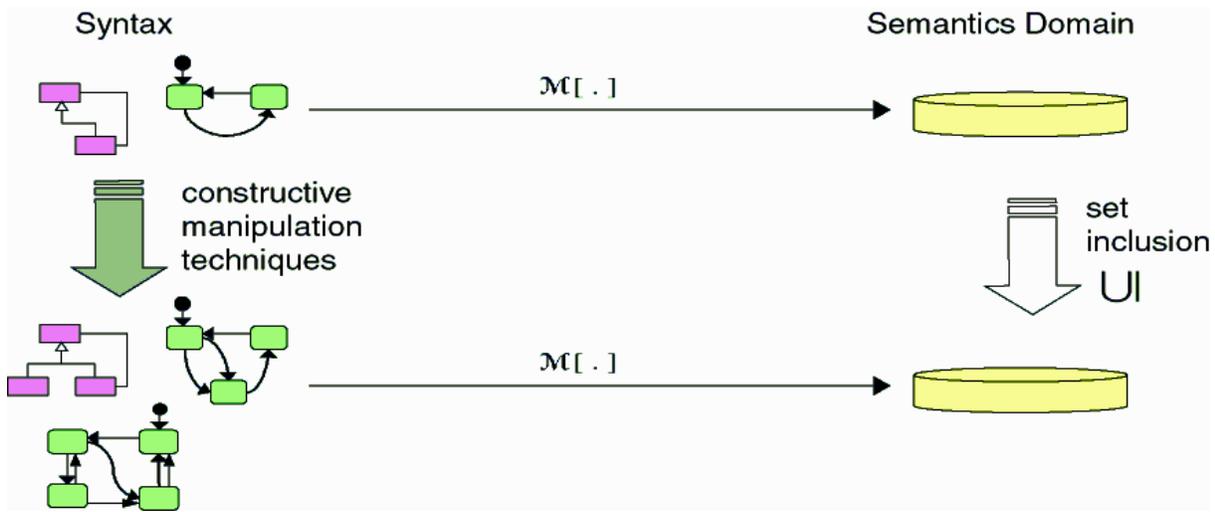

Figure 2: Commuting diagram for constructive manipulation techniques with proper semantics.

Please note as our example was a rather small one, we did not give a concrete graphical or textual representation except by illustration in Figure 1. A concrete representation would be necessary to make real use of a notation. Furthermore, if infinite sets of states, messages or an infinite set of transitions are of interest, an additional layer is needed to introduce finite representations of these infinite sets. This is e.g. in [11] done by using predicates to characterize equivalence classes of object states in one automaton state. The expansion of the predicates is part of the first level of semantics mapping and would lead here to an HMA as defined above.

## 5  Variants of Semantics Mappings

In Section 3, we have introduced the basic concepts for a semantics definition. They are based on the mapping $\mathcal{M}$ from new concepts of the language to define $\mathcal{L}$ into well known concepts of the semantics domain $\mathcal{S}$. These mappings come in many different variations that it currently seems not feasible to classify all of them. Instead we focus on several variants of these mappings and discuss their impacts. The following discussion can be regarded as a set of guidelines for the definition of semantics, also they do not help defining semantics of complex languages like programming languages. For this purpose one of the standard denotational semantics papers [30] [32] [33] [34] could be consulted.

### 5.1  Intended Audience of the Semantics Mapping

Very important is the intended audience of the semantics definition. Possible groups of readers are:
- the notation developer or the semantics definer,
- the tool vendor, or
- the user of a language.

If the semantics definition is made for the user of the notation then forget about formulae. This is at least true, if the user does not wish to have a formal definition but an intuitive appealing description of the purposes of the notation. As stated already, typical users will not be willing to understand the language $\mathcal{S}$ if it is given in an notation $\mathcal{N}_s$ that they aren't trained in. This would mean that first the notation $\mathcal{N}_s$ has to be understood, which itself is defined using another formal language, and so on. Even if the user has skills in formal methods or mathematics, there still might be resistance to learn $\mathcal{N}_s$. So as we do not have a semantics formalism that is commonly understood by a broad range of users (except perhaps mathematics) it is probably the best idea to use a natural language for explaining the notation and carefully describe the semantics.

The notation developer, who ideally should be the definer of the semantics, instead can cope with the notation $\mathcal{N}_s$ to gather insights for the specification language $\mathcal{L}$.

Tool vendors also may be exposed to a precise semantics, but it is probably better to give them detailed descriptions of what to handle if documents are iteratively changed. This includes simple rules for adding, removing and adapting elements of the notation, as e.g. in [21], or refinement and transformation calculi for them as given in [11], [15], or [16]. Strictly spoken, tool vendors are not interested in "what" a notation means, but in "how" its symbols can be modified, and "how" code can be generated from it. However, tool vendors should not forget that the latter two depend on the first.

The intended audience and their particular background of course has several impacts on the following topics to discuss. In our state machine example we mainly used the semantics definition for ourselves to precisely define the intended semantics and to prove certain interesting properties.

## 5.2 Degree of Formality

The degree of formality and precision of a semantics definition very much depends on the background of the definer and the intended audience. A semantics definition given in a formal notation is not necessarily precise (or let's say accurate), and a semantics given in pure English may be rather precise. Especially stylized English as suggested for the RM-ODP or diagrams as given in UML can be formal, provided there is a formal, commonly accepted semantics for them. On the one hand, however, there is a correlation between the preciseness of a definition and the language used. On the other hand, as discussed above, the degree of precision of a definition also depends on the background resp. knowledge of the user.

## 5.3 Abstraction and Information Loss

A document written in notation $\mathcal{L}$ contains a set of information pieces. This information tells the reader about the structure, the functionality, interaction patterns, collaborations, timing constraints, etc. of the components of a system. But such a document also contains lots of syntax related information, like layout, which typically does not influence the semantics. In between these two contrasts there are pieces of information that do contribute to the semantics of a document, but they should not explicitly shine up. In our case the state structure was not part of the semantics, although it contributed to it. Such information must be abstracted from in order to gain a useful semantics. Semantics in this case becomes more compact and also clearer.

In order to ensure that the semantics definition does not abstract from essential parts of the intention it is important at first hand to ensure that the intended information can be represented in the semantics domain $\mathcal{S}$. If we had decided that states are visible by other objects, we would have to incorporate the state set within the semantics, which in this case leads to a less abstract semantics. From a practical point of view it sometimes helps to split the semantics, by defining two mappings $\mathcal{M}_{beh}$ as above and $\mathcal{M}_{state}: \mathcal{HMA} \to \wp(S)$ to describe the state set.

Sometimes it is useful not to represent all essential information in the semantics, but to come up with a rather simple semantics definition especially useful for a certain aspect. For CSP there are quite a lot of different semantics available [18] that try to be as abstract as possible but still complete with respects to certain desired properties like e.g. compositionality. In our example, we decided that semantics consists of finite observations of behavior. We therefore cannot reason about liveness properties, as this requires infinite observations.

Quite the inverse happens if the semantics mapping is regarded as a transformation that adds information. This frequently occurs when the semantics domain $\mathcal{S}$ is itself a notation and the mapping is furthermore to be implemented. Typical examples are mappings of class diagrams to relationship models or cross-compilers from one programming language into another. There are lots of different choices to map new concepts into a language which is less rich in concepts. Performing these choices is equivalent with adding information as this also rules out unwanted implementations.

### 5.4 Compositional and Refinement-respecting Semantics

The problem discussed above of defining semantics as abstract as possible but still including all essential information can be captured more precisely when techniques for manipulation of the notation are given. These techniques usually fall in the two categories refinement and composition. Composition means that several documents of the language are composed resulting in a new document of the language. The binary versions of composition can be captured as:

$$.\oplus. : \mathcal{L} \times \mathcal{L} \to \mathcal{L}$$

A refinement transformation that modifies one document can be denoted as:

$$R : \mathcal{L} \to \mathcal{L}$$

Often these two operators come in parameterized versions, allowing various different flavors. E.g. composition of objects in a data-flow architecture can be sequential or parallel. Regarding our example, we have already mentioned several possible forms of refinement, however, if we wanted to introduce composition of HMA we should clarify more details about the communication of the underlying objects. There are different variants of object communication, but almost all lead to principal problems that do not allow us to describe the composition result itself as an HMA. This mainly stems from the problem of internal interactions of HMA that are not visible from outside, but can be influenced through sending new messages. This problem can be overcome by extending the language itself with a composition operation, which was e.g. done with StateCharts [19], where composition becomes a very simple operation: Take the two HMA, draw a box around them and separate the two HMA by drawing a line to indicate a so called "AND-state" that distributes control among subautomata. Please note, that this is not included within our HMA as we don't have AND-states.

It is important to ensure that the semantics of the composite can be deduced from the semantics of its parts. If this holds for all variants of compositions then the semantics definition is called compositional [20].

A quite similar approach could be useful to better understand the general need for semantic definitions that cope with refinement.

### 5.5 Algorithmic Mapping

For the primary purpose of giving the language $\mathcal{L}$ a meaning, neither the semantics domain $\mathcal{S}$ nor the mapping $\mathcal{M}$ need to be given in any implementation oriented style. However, in some cases the semantics definition can be used for a transformation by the user of $\mathcal{L}$. In our example this might be the case for the first part of the mapping, which can be easily implemented if the involved sets are finite. In this case not only the language $\mathcal{L}$ but also the semantics domain $\mathcal{S}$ needs a computer-based representation. The underlying notations $\mathcal{N}_\mathcal{S}$ and $\mathcal{N}_\mathcal{L}$ can then be regarded as some implementation data structure, depending on the programming language used for $\mathcal{N}_\mathcal{M}$. This works for transformations of Class Diagrams to Entity/Relationship models or cross-compilers between programming languages. These mappings, if given to the user, are not really semantics definitions, but syntactical transformations and should therefore have themselves a justification using an abstract semantics definition. Otherwise we end up like compilers in the early days of computer science: The semantics of the language was defined using the compiler, and the programmer just had to try.

There are example formalizations for diagrammatic languages $\mathcal{L}$ that essentially translate them into formulae of a logic language $\mathcal{N}_\mathcal{S}$. The idea was to give the user the possibility at hand to feed the generated formulae into a verifier such as Isabelle [22] and to have a precise tool at hand to verify properties of the defined model. Such an approach exhibits several problems. First, if you are not sure whether the generation algorithm is correct, the verified properties of the generated formulae might not hold for the original properties of the diagram. Second, if the generated formulae are modified, e.g. by enhancing them with certain properties, it is in general not possible to fully retranslate them into the diagrammatic language $\mathcal{L}$, as formulae usually have more expressive power. Third, most people accept diagrammatic languages, but very few like to see their intuitive diagrams vanish in a large set of generated formulae. This becomes even more of a problem, as usually the generated formulae are rather large and therefore hardly understandable.

There have been some ideas around that try to retranslate modified formulae, say by a mapping $\mathcal{F} : \mathcal{S} \rightarrow \mathcal{L}$. The mapping must not necessarily be complete, but perhaps may "lose" information. However, if such a situation arises, one should always be very eager to check out what the combination $(\mathcal{F} \circ \mathcal{M})$ does, as this ideally should be an identity. As these translations essentially work on the underlying notations, it can be doubted that a result is indeed the origin, but a much larger representation, which could (or should?) essentially describe the same systems. Quite the same is true for $(\mathcal{M} \circ \mathcal{F})$.

Another kind of mappings that is not quite algorithmic, but also intended to be available to the user is the semi-automatic translation. In this case the software developer is actively involved in the translation process by deciding which offered choice to take or which pieces of information to add. The transformation is such interactive and therefore merely a syntactical manipulation and not a semantics definition. Examples also come from database design, where e.g. the transformation of an association allows to choose from different implementations. So as opposed to above, not the transformation itself adds this extra information, but the user is required to provide it.

### 5.6 Choice of Underlying Notations

A rather technical but nevertheless important question is the choice of the appropriate languages to describe syntax and semantics. There are various relationships between the language and the semantics in question and the underlying notation used to describe them. In our example, we have used mathematics as the notation to define language, semantics domain and mapping. Other examples, especially the ones using Z often enhance the language $\mathcal{L}$ by concepts used in the semantics domain $\mathcal{S}$. Therefore the notations of both are not disjoint and the mapping becomes much easier. For example StateCharts can be enhanced with pre- and postconditions using Z and these conditions need therefore not be changed when mapped. However, this has the severe drawback that the mapping cannot be given explicitly within Z.

To clarify this, regard Z to be used either as semantics domain $\mathcal{S}$ or as the underlying notation $\mathcal{N}_\mathcal{S}$ to define the semantics domain. In the former case, the result of the mapping is a Z specification. Z itself is precisely defined using grammars, which serve as $\mathcal{N}_\mathcal{S}$. In the latter case Z is used to define the semantics domain. This can e.g. be relations between sequences of input and output messages and has nothing to do with Z.

It is possible to use Z for both, $\mathcal{S}$ and $\mathcal{N}_\mathcal{S}$. This would lead to a formal representation of the abstract syntax of Z in itself and has not been done so far. Interestingly, the developers of UML did exactly this to define their own notation, when they used UML for $\mathcal{L}$ and $\mathcal{N}_\mathcal{L}$.

If Z is used for $\mathcal{S}$ usually the mapping is not explicitly stated, but given through informal rules.

## *5.7 Choice of the Semantic Domain*

The choice of the semantic domain $\mathcal{S}$ of course depends on the topics mentioned earlier. One general approach at defining the semantics is to choose the underlying notation $\mathcal{N}_\mathcal{S}$ and do not talk explicitly about the semantic domain $\mathcal{S}$ itself. This is common, when choosing Z or a similar notation for $\mathcal{N}_\mathcal{S}$, as it then happens that $\mathcal{S}$ is implicitly given as the set of all valid Z specifications. But usually by far not every Z specification can really be generated by $\mathcal{M}$. If a mapping $\mathcal{M}$ is given, it might be a good idea to try to find some intrinsic properties, when examining the range of $\mathcal{M}$, denoted as $\mathcal{M}(\mathcal{L})$. This usually is a true subset of $\mathcal{S}$ and therefore can be regarded as the implicitly given description of the systems we are interested in.

Whenever a document of the language $\mathcal{L}$ is translated, the mapping $\mathcal{M}$ adds a lot of extra properties to describe certain situations. A good example for such a property is that if a class inherits from another one, it has at least the methods and the attributes of its ancestors. This property is completely independent of the actual set of classes, and also completely independent of the form of the notation, we use to describe these classes, but is an intrinsic property of the intended semantic domain. Thus it should be described within $\mathcal{S}$ and not within the mapping $\mathcal{M}$. This on the one hand allows to explicitly understand and reason about intrinsic properties of $\mathcal{S}$ without having to bother with a concrete syntax, and on the other hand allows to exchange or extend the notation $\mathcal{L}$ without modifying $\mathcal{S}$.

As there are extensions, like OOZE [23], or Z++ [24] of Z to incorporate object-oriented concepts like classes, it is quite a natural choice to use this extensions. The more concepts $\mathcal{L}$ and $\mathcal{S}$ have in common the easier the mapping becomes. In our examples we used this idea to decompose our semantics definition into several easier ones, e.g. one dealing with the reduction of state hierarchies only. One pitfall to avoid is the problem that different languages, like UML and OOZE do both have the notion of object and class in common, but they don't necessarily match in detail. For example in OOZE identity and especially identifiers are treated different from a procedural programming language. This can lead to an incorrect semantics mapping, which in best case is just a loss of information and therefore an incomplete formalization.

It is an intention of OOZE to study essential concepts of object-orientation in an abstract way. However, using OOZE as semantics domain may lead to the problem that a direct and easy mapping between similar concepts in $\mathcal{L}$ and $\mathcal{S}$ is inappropriate.

## 5.8 Construction of the Mapping

As we have shown in our example a semantics definition can be constructed in several steps. This is mainly used to reduce the complexity of the overall mapping. Two main techniques of decomposing a mapping (which in essence is a function) can be applied:

- Functional decomposition of $\mathcal{M}$.
- Split of the source language $\mathcal{L}$.

The first technique was applied in our example, where several steps have been taken. Chains of mappings are connected together by functional composition:

$$\mathcal{L}_0 \xrightarrow{\mathcal{M}1} \mathcal{L}_1 \xrightarrow{\mathcal{M}2} \quad ... \quad \xrightarrow{\mathcal{M}(n-1)} \mathcal{L}_{n-1} \xrightarrow{\mathcal{M}n} \mathcal{L}_n = \mathcal{S}$$

We have for example used this technique when defining semantics for HMA.

The second technique can be applied if a modeling technique like UML consists of a set of rather independent notations and that a semantics definition for them can be given independently from each other. We therefore split the source language $\mathcal{L}$ into several (not necessarily disjoint) sublanguages:

$$\mathcal{L} = \mathcal{L}^1 \cup \mathcal{L}^2 \cup ... \cup \mathcal{L}^n$$

and define a semantics for them in the form of

$$\mathcal{M}^i : \mathcal{L}^i \to \mathcal{S}$$

If the semantics domain is chosen carefully, namely if there are operations to integrate several elements of $\mathcal{S}$ into a single one, then the integration of the mappings $\mathcal{M}^i$ into one mapping $\mathcal{M}$ should be possible. A natural choice for an $\mathcal{S}$ allowing this is to use a set valued semantics and use intersection to integrate. Below we discuss set based semantics in greater detail.

The given techniques can be combined, e.g. by functionally decomposing the above $\mathcal{M}^i$ by first mapping $\mathcal{L}^i$ into an appropriate semantics $\mathcal{S}^i$ and then embedding the result in a general semantics domain $\mathcal{S}$ describing all facets of systems. In our example, we may add class diagrams describing the structure, give them a semantics on their own, and then embed both semantics in a more general semantics domain, capable of describing all interesting facets of systems.

As we can see from these composition techniques for semantics definitions, it to some extent lies in the eye of the beholder and also depends on the given context to regard one language as syntax and the other one as semantics. When using functional decomposition, the question what is syntax and what is semantics depends pretty much, which mapping we are focusing on. With respects to mapping $\mathcal{M}_k$ the language $\mathcal{L}_{k-1}$ is regarded as syntax and $\mathcal{L}_k$ as semantics domain. Please also note that some languages do apply to serve as syntactic notation much better, than others. A syntactic language must be finitely represented, otherwise there must be an additional layer that incorporates concepts to finitely represent infinite sets, relationships etc. A good example for that is the use of state predicates to group together a possibly infinite set of states an object can have into one automaton state.

## 5.9 Set-valued Semantics

When a (deterministic) programming language is considered, the semantics describes for each possible input exactly one possible result. Therefore there is exactly one system that implements the intended semantics of a program.

The intention of specification languages like the algebraic specification language Spectrum [8] or the model-oriented languages Z and VDM, is to describe not a single system, but a set of systems whose elements have the given properties. The semantics for them is given as sets of algebras (models).

Quite like these specification oriented approaches we found it extremely useful to use a set valued semantics. This allows us to describe interesting properties of our semantics without regarding any details of the syntax. Let us assume $\mathcal{Y}$ describes the important properties of the set of possible systems we are interested in. In our HMA example this was a tuple of message sequences (M* × M*). The semantics domain is then given as a powerset: $\mathcal{S} = \wp(\mathcal{Y})$. The semantics of a document $D \in \mathcal{L}$ is thus a subset of systems from $\mathcal{Y}$, which we expect is the set of all systems that obey the properties stated in document D: $\mathcal{M}(D) \subseteq \mathcal{Y}$.

We can then state important properties:

- Document D is consistent exactly if $\mathcal{M}(D) \neq \emptyset$, which means an implementation can be found. Otherwise some contradicting propositions must be present within D.
- The semantics of two documents $D_1$ and $D_2$ is given by $\mathcal{M}(D_1, D_2) = \mathcal{M}(D_1) \cap \mathcal{M}(D_2)$, which essentially resembles that only systems that obey both documents are in the semantics.
- Document $D_2$ refines $D_1$, exactly if $\mathcal{M}(D_1) \subseteq \mathcal{M}(D_2)$, which means that $D_1$ can be replaced by $D_2$ without loss of information.

The second property allows us easily to integrate the semantics definition of documents of different notations. Given several notations $\mathcal{L}^i$ (e.g. Class Diagrams, Sequence Diagrams, HMA) and mappings $\mathcal{M}^i : \mathcal{L}^i \to \wp(\mathcal{Y})$ then an integrated semantics of a set of such documents can be obtained using intersection, to an integrated semantics mapping $\mathcal{M}$.

As stated already it is an important task for the notation developer to provide context conditions to ensure consistency of documents. For the definition of correct refinement rules and composition operations another set of context conditions has to be found.

When an executable specification language is regarded, then there often are two kinds of semantics. Our HMA example as well as e.g. CafeOBJ [13] provide an operational semantics, which describes one possible system, and an algebraic one, which describes a set of systems. The operational semantics is a strict subset (typically containing only one element) of the algebraic semantics, thus selecting a canonical implementation. Such an implementation semantics is in some approaches also called minimal or initial. If the language $\mathcal{L}$ in question is regarded to be executable, then the approach of a set-based, loose specification semantics and an additional implementation semantics could be appropriate. We used such an approach in our example when defining $\mathcal{M}_i$ and $\mathcal{M}_s$ for HMA.

# 6 Future of Semantics for UML

UML as the unification and enhancement of previously given object-oriented methods is a great achievement. However, due to the lack of a more precise semantics there is lot of space for different interpretations for different users. This partly depends on the personal background but also on the interpretation the tool vendors use.

Furthermore, there are still deficiencies between the interrelationships of different notations that remain to be solved. One famous example in the history have been the data-flow diagrams of OMT [25], which for themselves have a precise semantics, but their correlation to other OMT notations never became clear, and finally have not been taken over to UML. The activity diagrams of UML might have a similar fate, as today it is still not clear what Activity Diagrams really are. UML regards them as specialization of State Diagrams, but they seem to have more in common with OMT's data flow diagrams.

Quite a few people do have what they think is a precise intuition about all the UML constructs. However, today we doubt that these intuitions coincide. Instead the personal background and education, as well as experience using the UML notation and tools strongly influences this intuition. As tools become more and more mature, and a standardization of the semantics of UML in detail did not take place so far, it will become increasingly difficult to find such a standardization that people and especially tool vendors can agree upon. This might be a clear drawback of UML in the future, especially if someone else comes up with a new approach with more precisely defined semantics.

In the semantics document of UML, there is an attempt for such a standardization of the semantics. But it cannot be expected that this is already complete and consistent. A standardization not necessarily needs to be done based on a formalization, but it surely will help at least the people involved to systematically find and correct flaws and at last give a precise semantics definition.

However, such a process may take time, as there are lots of details to agree upon. This e.g. starts with the choice of the underlying notation $\mathcal{N}_s$. One (today very optimistic) hope would be that the choice of the underlying notations does not influence the semantics of UML. Although the difficulty to describe the semantic domain $\mathcal{S}$ differs with the choice of $\mathcal{N}_s$, the semantic domain and also the mapping can essentially describe the same "semantics". So several different groups could try to apply different variants of semantics definition and try to find common insights and conclusions.

Whichever technique will be used to standardize the semantics of UML, developers must not be exposed to it. Someone who knows UML as the only notation will never accept any formalization based on a mapping to any other notation. So, precision as well as formalization is to this respect quite subjective. It depends on the viewpoint and background of the individual persons.

Today some evidence arises that UML will more and more be used not as a specification language but as a high level programming language. This has some advantages, as if the concepts of UML are executable, they can immediately be animated and tested, or the generated code even be used as implementation. Thus UML probably will have an implementation-oriented semantics describing this animation.

Originally UML was intended to serve as a specification language. But a specification is primarily intended to describe properties of systems that the system developers want to be valid, but to leave open other properties that are not clear already. Today this is partly achieved by having a semantics that is rather vague (and here we mean imprecise as opposed to not detailed). However, this is not an advantage, as the developer cannot fix this kind of impreciseness within UML, but can adapt the individual interpretation only. Furthermore, to get complete (and therefore executable) UML descriptions, often certain details have to be specified, which the developer does not yet know or wants to leave open to a later phase of development or even implementation. It is an intrinsic problem of executable languages that this kind of over-specification frequently occurs. Instead it would be of some help to have flexible concepts of under-specification to postpone detail decisions to situations, where the decisions can and must be made.

In our example, as well as in [26], where a formalization of UML is discussed, we overcame this problem by defining a set based semantic domain, thus allowing to describe sets of intended systems through documents. In these sets of systems there are distinguishable canonical elements, that can be used as implementation semantics. A sublanguage for UML could be restricted to a subset of the notations that enforces the documents are executable from a certain detailed level. The executable sublanguage would have an implementation semantics, selecting a canonical (minimal) system as implementation. The combination of these two semantics would enable us to make UML a more abstract and flexible language than it is today: a missing transition is interpreted as fully under-specification (everything allowed) and not with a default implementation (no effect), thus allowing to add transitions later in the development without changing the information present, but refining it. There is a also some evidence that even two kinds of semantics are not sufficient for a notation like the UML. Instead it might be necessary to distinguish the semantics for UML diagrams even between the analysis and the design phase.

# 7 Conclusion

There exists a tremendous amount of work that defines semantics for a variety of programming and specification languages. From that a lot of principles for defining denotational or axiomatic semantics emerged that can be applied for today's new diagrammatic languages and notations. These principles can clearly help defining a precise semantics for them, even if not applied in a formal manner. It was the aim of this paper to give an overview of the different variants of a general approach that can be regarded as some general guidelines for defining semantics.

We have argued that tool vendors are not precisely interested in "what" the semantics of a notation is, but mainly in "how" to maniulate the syntax. The situation might probably be even worse: tools vendors might not be interested in a standardized precise semantics at all, as this would also standardize parts of their tools and therefore would not allow a vendor to be "better" than the others in terms of having better modification techniques (code generation, etc.). However, it should be a vital demand of the users to enforce a greater standardization of the semantics, which in itself is only possible through a formalization of the notation, as this would greatly increase interoperability between tools. Currently, we will probably end up with the situation for early programming languages 30 years ago, where each compiler defined its own semantics.

The degree of formality and precision of a semantics definition very much depends on the background of the definer and the intended audience. The principles for a semantics definition discussed in this paper apply to the full range of semantics definitions, formal and informal ones. The benefits of a "formalization" mapping are less that it is "formal" afterwards, but more that the mapping of a new notation into a given formalism let become properties of the new notation become apparent that had been hidden before. This allows a deeper and better understanding of the new notations and eventually leads to an enhancement of the notation and its tools.

Perhaps it will be a good measurement for the precision and formality of UML, when the first papers appear that do not formalize UML, but use UML as the destination of a formalization, regarding UML as the well known notation $\mathcal{N}_g$.

### Acknowledgements

Thanks go to the many colleagues in Munich for many interesting discussions, and especially to Ingolf Krüger, as well as Haim Kilov, Peter Scholz and Oscar Slotosch for their comments on an earlier version of this document. I would also like to thank numerous people I had the chance to discuss with at recent conferences and workshops, and especially to the other members of the PreciseUML group.

This work was carried out within the SysLab project, funded by the Deutsche Forschungsgemeinschaft (DFG) under the Leibniz Preis Programm and Siemens Nixdorf.